\documentclass[11pt]{amsart}

\usepackage{latexsym,amsfonts,amssymb,epsfig,tabularx,amsthm,dsfont,enumerate}
\usepackage{amsfonts}
\usepackage{amscd}
\usepackage{color}
\usepackage{amsmath}
\usepackage{amsmath}
\usepackage{float}
\usepackage{amsthm}
\usepackage{amssymb}
\usepackage{mathrsfs}
\usepackage{amstext}
\usepackage{graphicx}
\usepackage{tikz}
\usepackage{verbatim}
\usepackage{framed}
\usepackage{hyperref}

\theoremstyle{definition}

\newtheorem{thm}{Theorem}

\newtheorem{lem}[thm]{Lemma}

\newtheorem{cor}[thm]{Corollary}

\newcommand{\N}{\mathbb{N}}

\newcommand{\Pn}{\mathcal P}
\newcommand{\B}{\mathcal B}

\newcommand{\wt}{\widetilde}
\newcommand{\eps}{\varepsilon}

\newcommand{\disp}{\,{\rm disp}}
\newcommand{\disc}{D}
\newcommand{\Bex}{\B_{\rm ex}}
\newcommand{\Bper}{\B_{\rm per}}

\newcommand{\bproof}{\begin{proof}}
\newcommand{\eproof}{\end{proof}}

\title{A lower bound for the dispersion on the torus}   

\date{\today}

\author{ 
Mario Ullrich
}

\begin{document}
 
\begin{abstract} 
We consider the volume of the largest axis-parallel box in the 
$d$-dimensional torus
that contains no point of a given point set $\Pn_n$ with $n$ elements. 
We prove that, for all natural numbers $d, n$ and every point set $\Pn_n$, 
this volume
is bounded from below by $\min\{1,d/n\}$. 
This implies the same lower bound for the discrepancy on the torus.
\end{abstract} 

\maketitle  


\section{Introduction}

The study of uniform distribution properties of $n$-element point sets $\Pn_n$ in the $d$-dimensional 
unit cube has attracted a lot of attention in past decades, in particular because of its 
strong connection to worst case errors of numerical integration using cubature rules, 
see e.g.~\cite{DP14,Niederreiter92,NW10}. 
There is a vast body of articles and books considering the problem of bounding the discrepancy 
of point sets. 
That is, given a probability space $(X,\mu)$ and a set $\B$ of measurable subsets of $X$, 
which we call \emph{ranges}, we want to find the maximal 
difference between the measure of a set $B\in\B$ and the empirical measure 
induced by the finite set $\Pn_n$, i.e. 
\[
\disc(\Pn_n,\B) \;:=\; \sup_{B\in\B}\, 
	\left| \frac{\#(\Pn_n\cap B)}{n} - \mu(B) \right|,
\]
where $\Pn_n\subset X$, $n\in\N$, with $\#\Pn_n=n$. 
In what follows we only consider $X=[0,1]^d$, $d\ge1$, and the 
Lebesgue measure $\mu$; we write $|B|:=\mu(B)$. 
The number 
$\disc(\Pn_n,\B)$ is called the \emph{discrepancy} of the point set $\Pn_n$ 
with respect to the ranges $\B$.
See e.g.~the monographs/surveys~\cite{DP10,DP14,DT97,Niederreiter92,No15,NW10} for 
the state of the art, open problems and further literature on this topic. 

Here, we are interested in lower bounds for this quantity that hold for every 
point set $\Pn_n$. 
In fact, we are going to bound the apparently smaller quantity
\[
\disp(\Pn_n,\B) \;:=\; \sup_{\substack{B\in\B\colon\\ \Pn_n\cap B=\varnothing}} |B|,
\]
which we call the \emph{dispersion} of the point set $\Pn_n$ with respect to the ranges $\B$. 
Clearly, this is a lower bound for the discrepancy. 

The notion of the dispersion was introduced by Hlawka~\cite{Hl76} 
as the radius of the largest empty ball (for a given metric). In this setting there are 
some applications including the approximation of extreme values (Niederreiter~\cite{Niederreiter83}) 
or stochastic optimization (Yakowitz et al.~\cite{YLV00}).
The present definition was introduced by Rote and Tichy~\cite{RT96} together with a treatment 
of its value for some specific point sets and ranges.
Only recently an application to the approximation of high-dimensional rank one tensors was 
discussed in Bachmayr et al.~\cite{BDDG14} and Novak and Rudolf~\cite{NR15}, where the ranges 
are all axis-parallel boxes in $[0,1]^d$. 
A polynomial-time algorithm for finding the largest empty axis-parallel box in dimension 2 
was considered by Naamad, Lee and Hsu~\cite{NLH84}.


Our main interest is the complexity of the problem of finding point sets 
with small dispersion/discrepancy; especially the dependence on the dimension. 
That is, given some $\eps>0$ and $d\in\N$, 
we want to know how many points are necessary 
to achieve 
$\disp(\Pn_n,\B)\le\eps$ or $\disc(\Pn_n,\B)\le\eps$ for some $\Pn_n\subset[0,1]^d$
and $\B\subset2^{[0,1]^d}$.
For this we define the inverse functions
\[
N_0(\eps,\B) \,:=\, \min\left\{n\colon \disp(\Pn,\B)\le\eps \text{ for some } 
	\Pn \text{ with } \#\Pn=n \right\}
\]
and 
\[
N(\eps,\B) \,:=\, \min\left\{n\colon \disc(\Pn,\B)\le\eps \text{ for some } 
	\Pn \text{ with } \#\Pn=n \right\}.
\]
We have $N_0(\eps,\B) \le N(\eps,\B)$ for every $\eps$, $\B$. 

For example, if $\B=\Bex^d$ is the set of all axis-parallel boxes contained in $[0,1]^d$, 
then it is easily seen that 
for every point set there exists an empty box with volume larger than $1/(n+1)$; 
simply split the cube in $n+1$ equal parts, one of which must be empty.
Moreover, it is known that 
with respect to the dependence on $n$ this estimate is asymptotically
optimal, i.e.~there exists a sequence of point sets 
$(\Pn_n)_{n\in\N}$ such that $\disp(\Pn_n,\Bex^d)\le C_d/n$ for some $C_d<\infty$, see 
e.g.~\cite{RT96}.\footnote{Note that 
for the discrepancy such an inequality cannot hold for any sequence of point sets, see Roth~\cite{Ro54}.}. 

However, if one considers increasing values of the dimension the situation is less clear: 
The best bounds to date are
\[
\frac{\log_2 d}{4(n+\log_2 d)} \;\le\; \inf_{\Pn\colon \#\Pn=n} \disp(\Pn,\Bex^d) 
\;\le\; \frac{C^d}{n}
\]
for some constant $C<\infty$, 
see Aistleitner et al.~\cite{AHR15} for the lower bound 
and Larcher~\cite{LaPC} for the upper bound.
For a proof of an super-exponential upper bound see also Rote and Tichy~\cite[Prop.~3.1]{RT96}. 
This can be rewritten as
\[
(1/4-\eps) \frac{\log_2 d}{\eps} \;\le\; N_0(\eps,\Bex^d) 
\;\le\; \frac{C^d}{\eps},
\]
Clearly, there is a huge difference in the behavior in $d$ for the upper and 
the lower bound. 

If we consider the discrepancy instead, then even the order in $\eps^{-1}$ differs in the upper and 
the lower bounds, i.e. for small enough $\eps\le\eps_0$ and all $d\in\N$ we have 
\[
c\, d\, \eps^{-1} \;\le\; N(\eps,\Bex^d) \;\le\; C\, d\, \eps^{-2}
\]
with some constants $0<c,C<\infty$.\footnote{If one considers only boxes that are anchored at 
the origin, i.e. the star-discrepancy, then one can choose $c=\eps_0=1/(32 e^2)\approx0.00423$~\cite{Hi04} 
and $C=100$~\cite{Ai11}.}
The lower bound is due to Hinrichs~\cite{Hi04} and the upper bound was proven by 
Heinrich et al.~\cite{HNWW01}. 
To narrow the gap in the $\eps$-behavior while keeping a polynomial behavior in $d$ is a 
long-standing open problem, see also~Novak and Wo\'zniakowski~\cite{NW10} for more 
results/problems in this area.

Nevertheless, for fixed, small $\eps>0$ the $d$-dependence of $N(\eps,\Bex^d)$ 
is known to be linear.
This motivates us to study the same problem for the dispersion.
Unfortunately, we were not able to this problem for the 
ranges $\Bex^d$. Instead, we consider the ``periodic'' version of this problem,  
i.e., we regard the unit cube as the torus and consider all axis-parallel boxes that respect 
this geometry. More precisely, we consider the ranges $\Bper^d$, see $\eqref{eq:Bper}$ 
and Figure~\ref{fig:Bper}, and we prove the following theorem.

\begin{thm}\label{thm:disp_per}
For every $n,d\in\N$ and every point set $\Pn_n\subset[0,1]^d$ with $\#\Pn_n=n$ we have
\[
\disp(\Pn_n, \Bper^d) \,\ge\, \min\{1,d/n\},
\]
or equivalently, 
\[
N_0(\eps,\Bper^d) \;\ge\; d/\eps \quad \text{ for }\quad 0<\eps<1.
\]
\end{thm}


\goodbreak

Clearly, this implies the following.

\begin{cor}\label{cor:disc_per}
For every $n,d\in\N$ and every point set $\Pn_n\subset[0,1]^d$ with $\#\Pn_n=n$ we have
\[
\disc(\Pn_n, \Bper^d) \,\ge\, \min\{1,d/n\}
\]
or equivalently, 
\[
N(\eps, \Bper^d) \,\ge\, d/\eps \quad \text{ for }\quad 0<\eps<1.
\]
\end{cor}

\medskip

As far as we know, the largest lower bound on the inverse of the periodic discrepancy 
that was known before is due to Hinrichs~\cite{Hi04} and states that 
\[
N(\eps, \Bper^d) \,\ge\, N(\eps, \B_*) 
\,\ge\, c\, d/\eps \quad \text{ for } 0<\eps<c,
\]
where $\B_*$ is the set of all axis parallel boxes that are anchored at the origin and 
$c>0$ can be chosen as $c=1/(32 e^2)\ge0.004229$. 
For the proof of this note that $\Bper^d\supset\B_*$.


\section{Preliminaries}

\begin{figure}[ht]
\begin{minipage}{0.48\textwidth}
\hspace{3mm}
\includegraphics[width=.8\columnwidth]{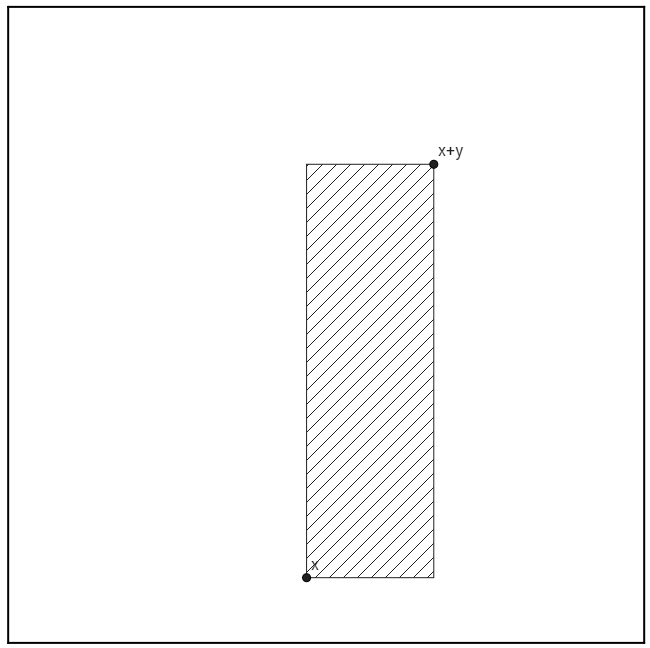}
\end{minipage}
\begin{minipage}{0.48\textwidth}
\includegraphics[width=.8\columnwidth]{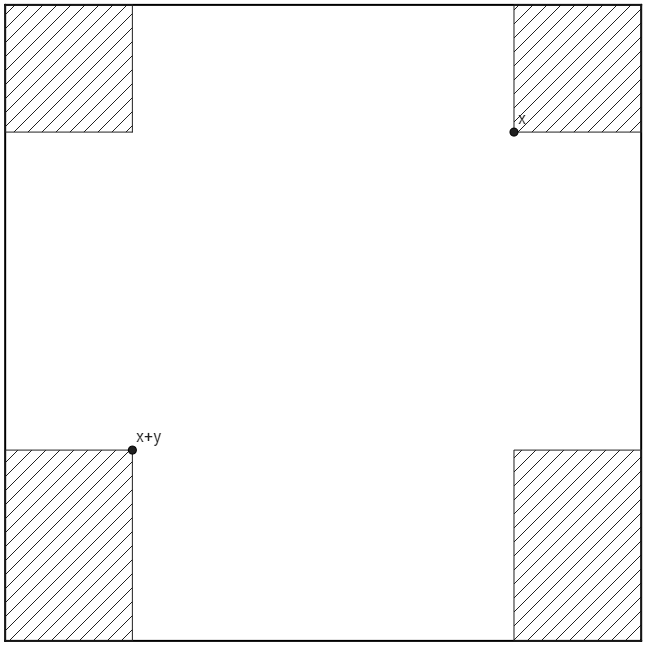}
\end{minipage}
\caption{Two sample test sets from $\Bper^2$}%
\label{fig:Bper}%
\end{figure}

For the dispersion on the torus, we consider 
ranges $B_1(x,y)\subset[0,1]^d$ of the form 
\begin{equation}\label{eq:boxes_per}
B_1(x,y) \;:=\; (0,y) \,+\, x  \mod 1
\end{equation}
for $x,y\in[0,1]^d$.
Note that $B_1(x,y)$ is simply $(x,x+y)$ iff $x+y\le 1$. In all other cases one has to 
respect the geometry of the torus, cf.~Figure~\ref{fig:Bper}.

We define the \emph{periodic ranges} by
\begin{equation} \label{eq:Bper}
\Bper^d \,:=\, \left\{B_1(x,y)\colon x,y\in[0,1]^d\right\}.
\end{equation}


\medskip

The main tool for the proof will be the following lemma, which provides a 
lower bound for the $d$-dimensional dispersion in terms of the
dispersion of certain projections of the point set. 

For a set $A\subset[0,1]^d$ we define the projections 
\begin{equation}\label{eq:proj}
A^{(k)} \,:=\, \bigl\{(x_1,\dots,x_k)\in[0,1]^k\colon (x_1,\dots,x_d)\in A \bigr\}, \qquad 
1\le k\le d,
\end{equation}
i.e.~we consider every element from $A$ without the last $d-k$ coordinates.
For a family of sets $\B\subset 2^{[0,1]^d}$ we define 
$\B^{(k)}=\{B^{(k)}\colon B\in\B\}\subset 2^{[0,1]^k}$.

Note that, if the ranges satisfy 
$\B\supset\B^{(k)}\times[0,1]^{d-k}=\{B\times[0,1]^{d-k}\colon B\in\B^{(k)}\}$ 
(as $\Bper^d$ or the set of all axis-parallel boxes), 
then it is obvious that
\[
\disp(\Pn_n,\B)
\;\ge\; \sup_{\substack{B\in\B^{(k)}\colon\\ \Pn_n\cap (B\times[0,1]^{d-k})=\varnothing}} |B|
\;=\; \sup_{\substack{B\in\B^{(k)}\colon\\ \Pn_n^{(k)}\cap B=\varnothing}} |B|
\;=\; \disp(\Pn_n^{(k)},\B^{(k)})
\]
for every point set $\Pn_n$. The same holds for the discrepancy.

However, such a bound is not sufficient to prove bounds on the dispersion 
that are growing with the dimension. Hence, we prove a refinement of this inequality.
In particular, we need the fact that we can forget about (at least) one point 
whenever we project to lower dimensions, loosing some specific constant. 
For this, we define 
\begin{equation}\label{eq:kappa}
\kappa_{\B}(A) \,:=\, \sup\left\{\kappa\ge0\colon 
	\forall \wt B\in\B^{(d-1)}\, \exists B\in\B\colon 
	\wt B=B^{(d-1)}, A\cap B=\varnothing \text{ and } |B| \ge \kappa |\wt B| \right\}
\end{equation}
for ranges $\B$ and a subset $A\subset[0,1]^d$.

\begin{lem}\label{lem:projection}
For every point set $\Pn\subset[0,1]^d$ and every $A\subset[0,1]^d$ we have 
\[
\disp(\Pn,\B) \,\ge\, \kappa_\B(A)\, 
\disp\left(\left(\Pn\setminus A\right)^{(d-1)},\, \B^{(d-1)}\right).
\]
\end{lem}


\bproof
By the definition of $\kappa_\B$ we obtain for every $\wt B\in\B^{(d-1)}$ 
and $A\subset[0,1]^d$ that
\[
\sup\left\{|B|\colon B\in\B \text{ with } B^{(d-1)}=\wt B 
	\text{ and } A\cap B=\varnothing\right\} 
\,\ge\, \kappa_\B(A)\,|\wt B|.
\]
Now take the supremum over all sets $\wt B$ with 
$(\Pn\setminus A)^{(d-1)}\cap\wt B=\varnothing$. Clearly, the right hand side 
then is $\kappa_\B(A)\,\disp\bigl((\Pn\setminus A)^{(d-1)},\, \B^{(d-1)}\bigr)$. 
The left hand side, after taking the supremum, is the supremum of the 
volumes $|B|$, where $B$ is such that $A\cap B=\varnothing$ and 
$(\Pn\setminus A)^{(d-1)}\cap B^{(d-1)}=\varnothing$. 
The second property implies that $\Pn\cap B\subset A$ and hence, 
by the first property, $\Pn\cap B=\varnothing$.
This shows that the left hand side is bounded from above by $\disp(\Pn,\B)$, 
which proves the statement.\\
\eproof


\section{Proof of Theorem~\ref{thm:disp_per}}

First we treat the case $n\le d$. 
Here, the advantage of the periodic test sets is most clearly observed. 
For $z\in[0,1]^d$ we consider the test boxes $B_1(z,1)$, cf.~\eqref{eq:boxes_per}, 
which consist of the whole cube without the $d$ hyperplanes $\{x\colon x_i=z_i\}$, 
$i=1,\dots,d$. 
Given an arbitrary point set $\Pn_n=\{x^{(1)},x^{(2)},\dots,x^{(n)}\}\subset[0,1]^d$ 
we define 
$z_i=x^{(i)}_i$ for $1\le i\le n$ and $z_i=x^{(n)}_i$ for $n+1\le i\le d$. 
Clearly, we obtain $\Pn_n\cap B_1(z,1)=\varnothing$ and $|B_1(z,1)|=1$. 
This proves $\disp(\Pn_n,\Bper^d)=1$ whenever $n\le d$.

\begin{figure}[ht]%
\includegraphics[width=0.4\columnwidth]{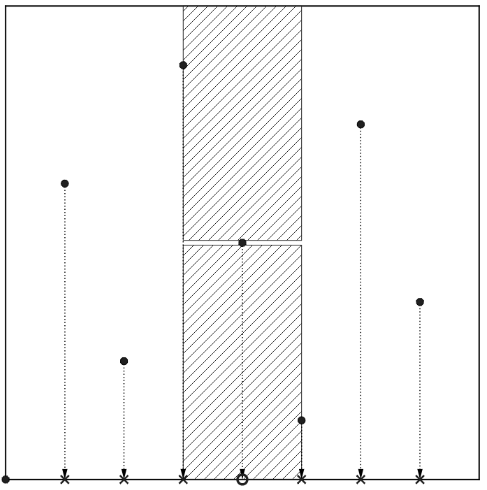}%
\vspace{-5mm}
\caption{The set $B\,=\,\wt B\times(0,1) + (0,\dots,0,t_d) \mod 1$}%
\label{disp_proj}%
\end{figure}

We now turn to the case $n>d$.
In this case we use Lemma~\ref{lem:projection}. 
Hence, we have to bound $\kappa(A):=\kappa_{\Bper^d}(A)$, 
cf.~\eqref{eq:kappa}, for specific $A\subset[0,1]^d$. In fact, we 
only need $A=\{t\}$ for some $t=(t_1,\dots,t_d)\in\Pn_n$.
Note that $(\Bper^d)^{(d-1)}$ is the set of all 
axis-parallel periodic boxes in $[0,1]^{d-1}$, i.e.~$(\Bper^d)^{(d-1)}=\Bper^{d-1}$. 
For every $\wt B\in\Bper^{d-1}$, we have that 
\[
B\,=\,\wt B\times(0,1) + (0,\dots,0,t_d) \mod 1 
\]
is an element of $\Bper^d$ that does not contain $t$, 
see Figure~\ref{disp_proj}.
Moreover, $|B|=|\wt B|$. 
This shows $\kappa(\{t\})=1$ for every $t\in[0,1]^d$ 
and, by Lemma~\ref{lem:projection},
\[
\disp(\Pn_n, \Bper^d) \,\ge\, 
\disp\left((\Pn_n\setminus\{t\})^{(d-1)}, \Bper^{d-1}\right).
\]
Iterating this procedure another $d-2$ times we obtain
\[
\disp(\Pn_n, \B_{\rm per}) \,\ge\, 
\disp\left((\Pn_n\setminus A)^{(1)}, \B_{\rm per}^{1}\right)
\]
for every $A\subset\Pn_n$ with $\#A=d-1$. 
Clearly, $\Bper^{1}$ is the set of all periodic intervals in $[0,1]$.
After taking the maximum over all 
$A$, the latter is the maximal length of a periodic interval that contains 
at most $d-1$ elements of $\Pn_n^{(1)}$. This is obviously bounded from below by 
$d/n$. This finishes the proof.\\
\qed

\bigskip


%



\goodbreak

\end{document}